\documentclass[11pt]{revtex4-2}

\usepackage{amsfonts}
\usepackage{amsmath}
\usepackage{color}
\usepackage[usenames,dvipsnames,svgnames,table]{xcolor}
\usepackage[colorlinks=true,
linkcolor=red,
urlcolor=blue,
citecolor=blue]{hyperref}
\begin{document}
	\title{Quantum Decomposition  Algorithm For Master Equations  of Stochastic Processes : The Damped Spin Case }

\author{ M.\ W.\ AlMasri} \address{Cybersecurity and  Systems Research Unit, ISI-USIM,\\ \small Bandar Baru Nilai, 71800 Nilai, Negeri
	Sembilan, Malaysia.} \author{ M.\ R.\ B.\ Wahiddin} \address{Cybersecurity and  Systems Research Unit, ISI-USIM,\\ \small Bandar Baru Nilai, 71800 Nilai, Negeri
	Sembilan, Malaysia.\\  Pusat Tamhidi, USIM, Bandar Baru Nilai, 71800 Nilai, Negeri
	Sembilan, Malaysia } 
	
	\begin{abstract}
	We introduce a quantum decomposition algorithm (QDA) that decomposes  the problem $\frac{\partial \rho}{\partial t}=\mathcal{L}\rho=\lambda \rho$ into a summation of eigenvalues times phase-space variables. One interesting feature of QDA stems from its ability to simulate damped spin systems by means of pure quantum harmonic oscillators adjusted with the eigenvalues of the original eigenvalue problem. We test the proposed  algorithm in the case of   undriven   qubit  with spontaneous emission and dephasing. 
\end{abstract}
\maketitle
	\section{Introduction}
	Stochastic processes are omnipresent in all areas of science such as game theory, finance,  probability theory, physics, chemistry (chemical kinetics), biophysics and others \cite{Blackwell,Van Kampen,Gardiner,Ross,Volovich,lecca}. The standard equation that describes the time evolution of stochastic processes is called the master equation. It assumes the form 
	\begin{eqnarray}\label{master0}
		\frac{\partial \vec{P}(t)}{\partial t}= M\cdot \vec{P}(t), 
	\end{eqnarray}
	where the vector $\vec{P}$ is a  column matrix in general and $M$ is (possibly time-dependent) transition matrix. The Chapman–Kolmogorov equation in probability theory, Fokker-Planck equation in quantum optics, Langevin equation for  Brownian motion (Wiener process) take the form of master equation \ref{master0}. In some cases $\vec{P}$ is not vector but a square matrix. For example $P$ is a $2\times 2$ matrix in  the quantum master equation for two-level systems with spontaneous emission. In this work, we propose new decomposition algorithm to express the time evolution of the square matrix say $\rho$ as a product of eigenvalue  times a phase-space variables i.e. $\frac{\partial \rho}{\partial t}\equiv \lambda \tilde{\rho}(z,w)$. In case of $2\times 2$ density matrix say $\sigma_{+}$,  $\frac{\partial \sigma_{+}}{\partial t}$  is just the eigenvalue times one-dimensional operators in the phase-space ( creation and annihilation operators of harmonic oscillators) not a square matrix. 
	\section{Solution of the quantum master equation in Damping basis }
	The time evolution of open quantum system in Markov approximation is determined by a {\it Born-Markov } quantum master equation of the form \cite{Kossakowski,Lindblad,Gorini,Davies,Zoller,Lendi,Breuer}
	\begin{eqnarray}\label{master}
		\frac{\partial \rho(t)}{\partial t}= \mathcal{L}\rho(t),
	\end{eqnarray}
Where the Liouville operator $\mathcal{L}$ is 
\begin{eqnarray}
	\mathcal{L}\rho=\frac{1}{i\hbar} [H_{S},\rho]+\sum_{k} \frac{\gamma_{k}}{2}\mathcal{D}[X_{k}]\rho
\end{eqnarray}
with set of bounded operators $X_{k}$ and real numbers $\gamma_{k}\geq 0$ and with $H_{S}$ being the system Hamiltonian . Here we define the superoperators of Lindblad form as 
\begin{eqnarray}\label{lindblad}
	\mathcal{D}[X]\rho= 2X\rho X^{\dagger}-X^{\dagger}X \rho- \rho X^{\dagger}X.  
\end{eqnarray}
For a time-independent Liouville operator, the solution of \ref{master} for initial density matrix $\rho(t=0)\equiv \rho(0)$ is of the form 
\begin{eqnarray}
	\rho(t)=e^{\mathcal{L}t}\rho(0)
\end{eqnarray}
In order to solve quantum master equations of the form \ref{master}, one could use the so-called {\it damping basis }approach introduced in \cite{damping} which enable us to express  \ref{master} as an eigenvalue problem 
\begin{eqnarray}
	\mathcal{L}\rho=\lambda\rho
\end{eqnarray}
Consider the space of linear operators $L(\mathcal{H})$ on a Hilbert space $\mathcal{H}$.  For $\rho_{1}$,$\rho_{2}\in L(\mathcal{H})$, the inner-product is 
\begin{equation}\label{inner}
(\rho_{1},\rho_{2})\equiv\; \ll \rho_{1}|\rho_{2}\gg\; \equiv\mathrm{Tr}\{\rho_{1}^{\dagger}\rho_{2}\}
\end{equation}
This definition of inner-product induces an  outer-product between two operators whose action is given by
\begin{eqnarray}
	[\rho_{1}\otimes\rho_{2}]\rho_{3}= \rho_{1}(\rho_{2},\rho_{3})= \mathrm{Tr}\{\rho_{2}^{\dagger}\rho_{3}\}\rho_{1}.
\end{eqnarray}
Since $\mathcal{L}$ is non-Hermitian operator with respect to the inner-product \ref{inner} i.e.
\begin{eqnarray}
	(\rho_{1},\mathcal{L} \rho_{2})=(\mathcal{L}^{\dagger}\rho_{1},\rho_{2})\neq (\mathcal{L}\rho_{1},\rho_{2})
\end{eqnarray}
it can be associated with left and right eigenvectors \cite{Stenholm}
\begin{eqnarray}
	\mathcal{L}|\phi_{k}\gg= \lambda_{k}|\phi_{k}\gg,\\
	\ll \mathcal{L}\varphi_{k}|=\ll \varphi| \lambda_{k},\\
	\mathcal{L}^{\dagger}|\varphi_{k}\gg=\lambda_{k}^{\star}|\varphi_{k}\gg. 
\end{eqnarray}
The spectral analysis of non-Hermitian operators is complicated in general. However, for sake of simplicity we assume the eigenvalues to be non-degenerate and both left and right eigenvectors to form an orthonormal  basis. That is 
\begin{eqnarray}
	\ll \varphi_{k^{\prime}}|\phi_{k}\gg=\delta_{k^{\prime},k}
\end{eqnarray}
 We can expand the density operator using the two sets of eigenvectors
 \begin{eqnarray}
 	|\rho(t)\gg= \sum_{k}r_{k}e^{\lambda_{k}t}|\phi_{k}\gg, \\
 	|\rho(t)\gg= \sum_{k} s_{k}e^{\lambda_{k}^{\star}t}|\varphi_{k}\gg, 
 \end{eqnarray}
here the coefficients are 
\begin{eqnarray}
	r_{k}e^{\lambda_{k}t}= \ll \phi_{k}|\rho(t)\gg, \\
	 s_{k}e^{\lambda_{k}^{\star}t}=\ll \varphi_{k}|\rho(t)\gg. 
\end{eqnarray}
From physical considerations, there must be a steady-state $\rho_{ss}$ corresponds to eigenvalue $\lambda_{ss}=0$. Any general quantum state should approach the steady-state preserving the probability interpretation of the density operator. Thus all eigenvalues should have non-positive real part $\mathrm{Re}\lambda_{k}\leq 0$ \cite{Stenholm}.\vskip 5mm
To summarize, we have the following eigenvalue problems for both right and left eigenvectors 
\begin{eqnarray}
	\mathcal{L}\rho_{\lambda_{k}}=\lambda_{k}\rho_{\lambda_{k}},\\
	\check{\rho}^{\dagger}_{\lambda_{k}}\mathcal{L}=\lambda_{k}\check{\rho}^{\dagger}_{\lambda_{k}}. 
\end{eqnarray}
The superoperator $\mathcal{D}$ defined in \ref{lindblad} acts on right eigenvectors. For left eigenvectors, we define $\check{\mathcal{D}}$ as
\begin{equation}
\check{\mathcal{D}}[X]\rho=2X^{\dagger}\rho X- X^{\dagger}X\rho-\rho X^{\dagger}X .
\end{equation}
For sake of simplicity, we assume the eigenvalues $\{\lambda_{k}\}$  to be non-degenerate. In this case, we can write 
\begin{eqnarray}
	\mathrm{Tr}\{\check{\rho}_{\lambda_{k}^{\prime}}^{\dagger} \mathcal{L}\rho_{\lambda_{k}}\}-\mathrm{Tr}\{\check{\rho}_{\lambda_{k}}^{\dagger}\mathcal{L}\rho_{\lambda_{k^{\prime}}}\}\\ \nonumber=\left(\lambda_{k}-\lambda_{k^{\prime}}\right)\mathrm{Tr}\{\check{\rho}_{\lambda_{k}}^{\dagger}\rho_{\lambda_{k^{\prime}}}\}=0
\end{eqnarray} 
which directly shows the left and right eigenvectors can be normalized according to the orthogonality relation 
\begin{eqnarray}
	\mathrm{Tr}\{\check{\rho}^{\dagger}_{\lambda_{k}}\rho_{\lambda_{k^{\prime}}}\}=\delta_{k^{\prime},k}.
\end{eqnarray}
In case the left and right eigenvectors form complete basis, the clouser relation reads
\begin{eqnarray}
	\sum_{k}\check{\rho}_{\lambda_{k}}\otimes\rho_{\lambda_{k}}=1
\end{eqnarray}
To put these ideas further into practical context, we consider a two-level system with excited state $|e\rangle$ and ground state $|g\rangle$\cite{damping,Stenholm}.  We denote the density matrix by $\rho$ and the transition frequency between the two levels by $\omega$. The free Hamiltonian for vanishing ground-state energy is 
\begin{equation}
	H=\hbar\omega \sigma_{e}
\end{equation}
The time-evolution of such an atom undergo spontaneous emission at rate $\Gamma$ and pure dephasing at rate $\Gamma^{\star}$ is described by Born-Markov master equation with Liouville operator equals to 
\begin{equation}
	\mathcal{L}\rho= -i\omega[\sigma_{e},\rho]+ \frac{\Gamma}{2}\mathcal{D}[\sigma_{-}]\rho+ \frac{\Gamma^{\star}}{2} \mathcal{D}[\sigma_{e}]\rho,
\end{equation}
where $\sigma_{-}=|g\rangle\langle e|$ is the lowering operator and its Hermitian conjugate gives the raising operator $\sigma_{+}=|e\rangle \langle g|$.  The steady state is $\rho_{ss}=|g\rangle\langle g|$ and corresponds to zero eigenvalue $\lambda=0$. The atomic coherence operators $\rho_{\pm}=\sigma_{\pm}$ correspond to $\lambda_{\pm}= \mp i\omega- \tilde{\Gamma}/2$ where $\tilde{\Gamma}=\Gamma+\Gamma^{\star}$. Finally for the density operator $\rho_{\rightarrow}=|e\rangle \langle e| -|g\rangle\langle g|$ the eigenvalue is $\lambda_{\rightarrow}=-\Gamma$.
\section{The  Algorithm}
Spin operators can be represented in terms of uncoupled harmonic oscillators using the Jordan-Schwinger map \cite{Jordan,Schwinger}
\begin{eqnarray}
	\sigma_{0}=a^{\dagger}a+b^{\dagger}b,\\
	\sigma_{1}= a^{\dagger}b+b^{\dagger}a , \\
	\sigma_{2}= -i(a^{\dagger}b-b^{\dagger}a), \\
	\sigma_{3}= a^{\dagger}a-b^{\dagger}b, 
\end{eqnarray}
where the canonical commutators are $[a,a^{\dagger}]=[b,b^{\dagger}]=1$ and $[a,b]=[a^{\dagger},b]=[a,b^{\dagger}]=[a^{\dagger},b^{\dagger}]=0$. In Schwinger oscillator model of angular momentum, the quantum angular momentum operators are defined as $J_{i}=\frac{\hbar}{2}\sigma_{i}$ where $i=1,2,3$ and the total number operator is $N=a^{\dagger}a+b^{\dagger}b$. In this model, the angular momentum eigenstates are  $|n_{a},n_{b}\rangle= \frac{a^{\dagger}}{\sqrt{n_{a}!}}\frac{b^{\dagger}}{\sqrt{n_{b}!}}|0,0\rangle$, where the quantum number $j$ can be identified as $j=\frac{n{a}+n_{b}}{2}$ and $m=\frac{n{a}-n_{b}}{2}$ running from $-j$ to $j$ into integer steps\cite{Sakurai}.
\vskip 5mm 
Now we solve  the problem of undriven qubit under spontaneous emission and dephasing using the Jordan-Schwinger map. We find the following relations
 \begin{eqnarray}
 	\mathcal{L}(\tilde{\rho_{ss}})= 0,\label{1}\\ 
 	\mathcal{L}(\tilde{\rho}_{+})=\left(-i\omega-\tilde{\Gamma}/2\right) \tilde{\rho}_{+},\label{2}\\
 	\mathcal{L}(\tilde{\rho}_{-})=\left(+i\omega-\tilde{\Gamma}/2\right) \tilde{\rho}_{-},\label{3} \\
 	\mathcal{L}(\tilde{\rho}_{\rightarrow})=-\Gamma \tilde{\rho}_{\rightarrow}\label{4}.
 \end{eqnarray}
where $\tilde{\rho_{ss}}=b^{\dagger} b, \tilde{\rho}_{+}=a^{\dagger}b/2, \tilde{\rho}_{-}=b^{\dagger}a/2, \tilde{\rho}_{\rightarrow}=a^{\dagger}a-b^{\dagger}b$. Now if we compare the last relations with previous section, we find that both pictures posses the same eigenvalues as expected but now the density operators are not square matrices but rather operators in the phase space. 
 \vskip 5mm
{\bf Theorem 3.1:} {\it Let $F,G,H\in \mathbb{R}$ and let  $\mathcal{L}\rho=\frac{\partial \rho}{\partial t}=\begin{bmatrix}
F && G-iH\\
G+iH&&  -F
\end{bmatrix}$, then the  decomposition of  problem $\mathcal{L}\rho=\lambda\rho$ into summation of eigenvalues $\lambda=\sum_{i} \lambda_{i}$ times functions of the operators $a$, $b$ and their Hermitian  conjugates always exist using the  Jordan-Schwinger map defined by the relations}
\begin{eqnarray}
\tilde{\sigma}_{1}=G\sigma_{1},\\
\tilde{\sigma}_{2}=H\sigma_{2},\\
\tilde{\sigma}_{3}=F\sigma_{3}. 
\end{eqnarray}
{\it Proof:}\\ Consider the ansatz $\rho_{+}= \tilde{\sigma}_{+}$. Clearly it is always possible to represent $\tilde{\sigma}_{+}=\frac{\tilde{\sigma}_{1}+i\tilde{\sigma_{2}}}{2}$ as $GH a^{\dagger}b/2$ using the transformation mentioned above. Thus it is clear that  $\mathcal{L}\rho_{+}$ gives the same eigenvalue as $\mathcal{L}(GHa^{\dagger}b/2)$. The same conclusion applied for all other possible ansatzs. \vskip 5mm 
{\bf Lemma 3.1: }{\it The previous theorem applies in the case of arbitrary  square matrix $\rho$.  }\\
{\it Proof:}\\
Since Pauli matrices form an orthonormal basis, it is always possible to write any square matrix $A$ in term of Pauli matrices times the entries of $A$. 
\vskip 5mm
{\bf Lemma 3.2:} {\it The generalization of  theorem 3.1 can be done in a straightforward manner using the suitable Jordan-Schwinger map for higher spin representation.  }\vskip 5mm 
It is important to mention that we have considered pure  harmonic oscillators in the decomposition process. These are non-damping and not in connection with any bath. The eigenvalue problem for damped harmonic oscillators can be found in \cite{damping}. \vskip 5mm 

Now in order to put these results into more mathematical friendly approach, we may use the well-known Bargmann representation to express the creation and annihilation operators by complex variables in the Bargmann space. This approach was initiated by Fock observation that the commutator $[\frac{\partial }{\partial z},z]=1$ is similar to  the canonical commutation relation $[a,a^{\dagger}]=1$\cite{Fock}. The rigorous treatment of the Hilbert space of analytic functions with Gaussian measure was done by Segal and Bargmann in \cite{Bargmann,Segal}.  \vskip 5mm
{\bf Definition 3.1: }{\it The Bargmann or Segal-Bargmann spaces $\mathcal{H}L^{2}(\mathbb{C}^{n},\mu)$	 are spaces 	of the holomorphic functions with Gaussian integration measure $\mu=(\pi)^{-n}e^{-|z|^{2}}$
	and  inner-product of the form \cite{Bargmann,Segal,Perelomov,Folland,Almasri0}}
\begin{eqnarray}
	\langle f|g\rangle_{\mu}=(\pi)^{-n}\int_{\mathbb{C}^{n}}\overline{f}(z)g (z)e^{-|z|^{2}}dz, 
\end{eqnarray}
Where $|z|^{2}=|z_{1}|^{2}+\dots +|z_{n}|^{2}$. Moreover, the monomials $\{z^{n}/\sqrt{n!}\}$ form an orthonormal basis in the Bargmann space since  

\begin{eqnarray}
	\int \frac{dz d\overline{z}}{\pi} e^{-|z|^{2}} \overline{z}^{n}z^{m}=n!\delta_{nm}.
\end{eqnarray}
The Jordan-Schwinger map in the two-dimensional Bargmann space $\mathcal{H}L^{2}(\mathbb{C}^{2},\mu)$ can be defined as \cite{Almasri}
\begin{eqnarray}
	\sigma_{0}=z\frac{\partial}{\partial z}+ w\frac{\partial }{\partial w}, \\
	\sigma_{1}=z\frac{\partial }{\partial w}+w\frac{\partial}{\partial z}, \\
	\sigma_{2}=-i\left(z\frac{\partial }{\partial w}-w\frac{\partial }{\partial z}\right), \\
	\sigma_{3} = z\frac{\partial }{\partial z}-w\frac{\partial }{\partial w}. 
\end{eqnarray}
In this case, the eigenstates are $f(z,w)=\frac{z^{n_{a}}w^{n_{b}}}{\sqrt{n_{a}!n_{b}!}}$. The equations \ref{1}, \ref{2},\ref{3},\ref{4} can be re-phrased substituting $a\rightarrow \frac{\partial }{\partial z}$, $a^{\dagger}=z$, $b=\frac{\partial }{\partial w}$ and $b^{\dagger}=w$. In this way we define the density operators as a function of these phase space coordinates i.e. $\tilde{\rho}(z,w)$.\vskip 5mm
{\bf The Quantum Decomposition Algorithm (QDA): }\vskip 5mm
 1. Determine the quantum master equation for qubits or multi-level systems: 
 \begin{eqnarray}
	\mathcal{L}\rho=\frac{1}{i\hbar} [H_{S},\rho]+\sum_{k} \frac{\gamma_{k}}{2}\mathcal{D}[X_{k}]\rho
\end{eqnarray}\\ 
2. Write the ansatz $\rho$ as a function of the operators $a$,$b$ and their Hermitian conjugates using Jordan-Schwinger map i.e. $\tilde{\rho}(a,b,a^{\dagger},b^{\dagger})$. \\
3- Solve the eigenvalue problem $\mathcal{L}\tilde{\rho}=\lambda \tilde{\rho}$\\
4- Repeat the procedure for all possible ansatzs and compute the eigenvalues. \vskip 5mm 

To conclude, we have proposed a quantum decomposition algorithm (QDA) that decomposes the quantum master equation of qubits and any multi-level system under various damping mechanisms into eigenvalue problem of the form $\mathcal{L}\tilde{\rho}=\frac{\partial \tilde{\rho}}{\partial t}=\lambda \tilde{\rho}$ where $\tilde{\rho}$ are density operators written with respect to the phase-space variables. The main advantage of the proposed algorithm is the possibility of simulating and solving the quantum master equation of any multi-level system  by means of two uncoupled harmonic oscillators only. Moreover, it helps in finding the eigenvalues of damped spins. One immediate application of the proposed algorithm appears in the study  of  the optimal population transfer in a dissipative three-level $\Lambda$ system since the  basis composed of the dark-bright states and the intermediate lossy state facilitates the solution of the corresponding optimal control problem \cite{optimal}.  Finally, it would be great   to augment the QDA into other algorithms such as Grover,  Quantum Approximate Optimization Algorithm (QAOA) and others to decompose the qubit with possible different dampings \cite{Deutsch,oracle,Grover,Shor,Nilesen,Farhi,Fawzi,Montanaro} and also to experimentally test QDA in the NISQ era\cite{Preskill}. Moreover, the same algorithm can be used in the case of any master equation of stochastic processes. For example,  in the study of  a multi-phase truncated  Lévy flights which has slow convergence rate to Gaussian process\cite{Stanley}. 

\vskip 5mm

{\bf Acknowledgment:} We are grateful to USIM for financial support.

\end{document}